# Metal Oxide Resistive Memory using Graphene Edge Electrode

Seunghyun Lee[1*§], Joon Sohn[1*], Zizhen Jiang[1], Hong-Yu Chen[1,2], H. -S. Philip Wong[1]

[1]*Department of Electrical Engineering and Stanford SystemX Alliance, Stanford University, Stanford, California 94305, USA.*

[2]*Present affiliation: Memory Strategy Group, SanDisk Corporation, 951 SanDisk Drive, Milpitas, CA 95035, USA*

* These authors contributed equally to this work.
§ Corresponding author. Electronic mail: seansl@stanford.edu

**The emerging paradigm of "abundant-data" computing requires real-time analytics on enormous quantities of data collected by a mushrooming network of sensors. Today's computing technology, however, cannot scale to satisfy such big data applications with the required throughput and energy efficiency. The next technology frontier will be monolithically integrated chips with 3-dimensionally interleaved memory and logic for unprecedented data bandwidth with reduced energy consumption. In this work, we exploit the atomically thin nature of the graphene edge to assemble a resistive memory (~3Å thick) stacked in a vertical 3D structure. We report some of the lowest power and energy consumption among the emerging non-volatile memories due to an extremely thin electrode with unique properties, low programming voltages, and low current. Circuit analysis of the 3D architecture using experimentally measured device properties show higher storage potential for graphene devices compared that of metal based devices.**

## Introduction

The rapid adoption of non-volatile memory technology such as Flash[1] has enabled a revolution in today's mobile computing. To date, the ever-increasing demand for higher density has so far been met through the development of multi-level storage cells and smart peripheral control circuitry that hides the inadequacies and imperfections of the memory cell[1]. However, the diminishing amount of stored charges and the increase in bit error rates that accompany feature-size scaling impose significant challenges for the future[2]. Further gains in memory performance and device density will require new breakthroughs in both atomic-scale technology and bit-cost-effective 3D device architectures[2,3].

Resistive random access memories (RRAM) based on metal oxide have shown considerable promise as a possible successor to Flash because of better endurance, retention, speed, lower programming voltages, and a higher device density[3-5]. These devices also use material sets and fabrication temperatures that are compatible with today's silicon technology[3,4], and offer the opportunity for future monolithic three-dimensional integration with logic computation units.

Graphene, an atomically thin crystal lattice of carbon atoms, is known for its unique electronic properties[6]. Both graphene and graphene oxides have been used in various memory devices, including RRAM[7-11], ferroelectric memory[12], and Flash memories[13] as electrodes and oxides.

In this work, the atomically thin (~3Å thick) edge of monolayer graphene was actively used as a SET electrode to form an atomically thin memory structure. We investigate the low energy consumption and the stacking potential of the device in a 3D architecture that is amenable to large scale manufacturing.

# Results

## Device structure in a 3D vertical cross-point architecture

Two layers of graphene RRAMs (GS-RRAM, GS stands for graphene SET electrode) were stacked to build a 3D vertical cross-point architecture as illustrated in Figs. 1a and b. In the figures, the TiN electrode, the $HfO_x$ layer, and the graphene electrode are depicted in yellow, green, and black, respectively. The fabrication process explained in Supplementary Figure 1 and 2. A transmission electron microscope (TEM) image of the device's cross-section is presented in Figs. 1c-e. The graphene edge contacting the memory element ($HfO_x$) is highlighted in red. We also fabricated RRAMs based on platinum electrodes (Pt-RRAM) as control devices. The Pt-RRAM (Figs. 1f, g) which was reported previously[14,15] has the same 3D structure as the GS-RRAM.

Such 3D architectures are part of an ongoing drive in the research community to adopt a bit-cost-effective architecture[1,2,14-17] with storage densities surpassing that of Flash technology (Supplementary Note 1). From past experimental results[16], the density of a 3D vertical RRAM array is known to be mainly limited by the sheet resistance and the layer thickness of the plane electrode, and not so much by the lithographic half-pitch, as it is in 2D architectures. This is due to the limitation of the pillar electrode resistance and the non-vertical etching angle resulting from trench etching through metal planes[16]. Graphene's sheet resistance per thickness (125 Ω per square at a monolayer thickness of 3Å when doped[18]) is significantly lower than that of any metal. All metal films are known to exhibit a steep exponential increase in sheet resistance as the thickness falls below 5 nm[15]. Graphene is also significantly easier to etch vertically than metal during pillar formation. Using a well-accepted reliability projection[15]—assuming programming voltage of 3V, $SiO_2$ thickness of 6 nm, half-pitch of 22 nm, and 1° of etch angle improvement —a maximum of 200 stacks will be possible for graphene RRAM as compared to the 60 stacks possible with conventional bulk-metal-based 3D RRAM (Supplementary Table 1).

In both of our RRAM structures, the conductive filaments of oxygen vacancies form at the oxide ($HfO_x$) similar to conventional metal oxide resistive memories. The number and the size of the conducting filament paths determine the two resistance states of the RRAM: the high resistance state (HRS) and the low resistance state (LRS). In the Pt-RRAM structure (Fig. 1g), TiN is used as the SET electrode as in most conventional devices with TiN-oxide-Pt structures [4,14]. In the GS-RRAM structure (Figs. 1a-e), however, the graphene electrode is used as the SET electrode to store (SET) and release (RESET) the oxygen ions during the programming process. This is fundamentally different from our previous work[19] on graphene RRAM where the TiN electrode was the SET electrode. The application of graphene as the SET electrode led to power consumption 120 times lower in this work compared to the previous work [19].

### The device characteristics of GS-RRAM

A comparison of the typical SET/RESET switching cycle of the GS-RRAM and the Pt-RRAM is shown in Fig. 2a (inset: magnified view of GS-RRAM plot). The SET programming is achieved by applying a positive voltage to the TiN electrode in the Pt-RRAM and a negative voltage to the TiN electrode in the GS-RRAM. The SET/RESET voltage and the RESET current distribution of GS-RRAM and Pt-RRAM after 50 cycles of switching are shown in Figs. 2b and 2c (the values for Pt-RRAM are in agreement with the references[14,15]). Importantly, the SET/RESET voltages and the RESET currents of GS-RRAM are considerably lower than those of Pt-RRAM. The resistance distributions of both the HRS and the LRS states at 0.1 V bias after 50 cycles for both devices are shown in Fig. 2d. Even with such low programming voltages and current, the memory window is larger for GS-RRAM compared to Pt-RRAM (Fig. 2d).

The power consumption of an RRAM cell is given by the product of the programming voltages and the currents[4]. Due to such low SET/RESET voltages and currents, the power consumption of the GS-RRAM is 300 times lower than that of the Pt-RRAM (Fig. 2e). In fact, the power consumption of the GS-RRAM is one of the lowest compared to recent reports on low power RRAMs (Supplementary Figure 3). From the pulse-mode endurance test with 500 ns width pulse (see Methods), the switching energy (switching voltage × current × pulse width = 0.2V × 2.3μA × 500 ns) was found to be around 230 fJ. We compared this value with the values of other emerging non-volatile memories, including RRAM, conductive bridge RAM (CBRAM), phase change RAM (PCRAM), and magnetic RAM (MRAM) in Supplementary Figure 4, and found the energy consumption to be comparable to the lowest known values.

**The oxygen ion migration and Raman imaging**

The mechanism behind the low power/energy consumption can only be explained by first understanding the oxygen ion migration during the switching process. Figures 3b and 3c illustrate the different ways the oxygen ions move and form conductive filaments during the programming process of the Pt-RRAM and the GS-RRAM. For Pt-RRAM, the TiN is the SET electrode and the conducting filaments in the oxide are formed via oxygen migration from $HfO_x$ to the TiN electrode (Fig. 3b) [4,14].

In a GS-RRAM, however, a negative voltage is applied to the TiN electrode during the SET process, and the oxygen ions move toward the graphene (Fig. 3c). Unlike in conventional metal, there will be an electrical potential gradient in graphene since graphene is relatively more resistive (~6kΩ per square) than a common metal. Hence, the oxygen ions will not accumulate at the edge but will migrate horizontally in the graphene and the oxide interface. In our previous work[20], we have shown how oxygen ions migrate on graphene during the programming process of the RRAM cell by employing Raman spectroscopy (Supplementary Note 2). In this work, the oxygen ion movement was also confirmed by monitoring oxygen dopants in graphene using Raman spectroscopy (Figs. 3d-h , also see Methods). One of the most pronounced indicators of dopants in graphene is the reduction of 2D peak intensity in a Raman spectrum[20,21]. In Fig. 3d, a typical change in the 2D peak (2670 $cm^{-1}$) intensity is observed for HRS → LRS → HRS transition. During the SET process (i.e. HRS → LRS), oxygen ions are inserted into the graphene, doping the film. Consequently, a decrease in the 2D peak intensity is observed. During the RESET process, (i.e. LRS→ HRS) oxygen ions are pushed back into $HfO_x$ from the graphene film. This results in an increase in 2D peak intensity. The Raman peak intensity of silicon (520 $cm^{-1}$) and the baseline are plotted in parallel to ensure that the references have not changed during measurement (see Methods).

The spatially resolved Raman spectroscopy results for the change in 2D peak intensity during the HRS → LRS → HRS transition are shown in Figs. 3f, 3g, and 3h, respectively. The blue square in Fig. 3e indicates the Raman-mapped region in the actual device. As the device is switched from HRS to LRS via the SET process, the change in the 2D peak intensity can be readily observed by the contrast difference. The statistical distributions of the changes in 2D peak intensity are also shown as histograms. Noticeable changes in the median values and the standard deviation of the 2D peak intensity are observed as the oxygen ions are inserted into and pushed back from the graphene film. This oxygen migration in graphene is also known to be aided by the Joule heating generated during the SET/RESET event[20,22]. Experimental studies also suggest that oxygen can be highly mobile in graphene[20,22] and can be used as an oxygen capturing layer[20,23]. As indicated in the literature[20], the oxygen may form a covalent bond with the broken bonds of graphene after the SET process, and the process is reversed during the RESET process (Supplementary Figure 5).

**The working mechanism**

The GS-RRAM offers significantly lower power consumption compared to Pt-RRAM due to three factors: low SET compliance current (Fig. 2a), low RESET current (Fig. 2c), and low programming voltages (Fig. 2b). The Pt-RRAM cannot be operated with such low currents or voltages, and shows severe degradation of the memory window when it is programmed with a lower compliance current (Supplementary Figure 6).

The low SET compliance current in GS-RRAM is possible due to a more resistive HRS and a larger memory window (Fig. 2d) compared to Pt-RRAM. Since the magnitude of the RESET current is directly proportional to the SET compliance current [4], the low RESET current is also related to these two factors. A systematic breakdown of the resistance components is necessary to understand the differences in LRS/HRS of the two devices (Fig. 4a). Three factors may contribute to the increased resistance of HRS in GS-RRAM compared to Pt-RRAM: the access series resistance $R_{series}$ from the graphene sheet compared to the Pt sheet, the difference of the TiN/oxide ($R_{int,TiN}$) and graphene/oxide($R_{int,G}$) interface, and the different sizes (Figs. 3b and 3c) of filamentary conduction paths in $HfO_x$ ($R_{filament,Pt}$ and $R_{filament,G}$). From a transmission line measurement (Supplementary Figure 7 and 8), we found that compared to Pt the additional sheet resistance and the contact resistance of graphene contributed little to the total resistance of HRS. On the other hand, the filamentary resistance and the interfacial resistance between materials (graphene, Pt, or TiN to $HfO_x$) dominated the total resistance change during the SET/RESET process.

It is known that in an RRAM structure, the resistance of HRS increases as the inverse of the cell area, roughly following Ohm's law[4]. Specifically, the higher HRS of the GS-RRAM compared to Pt-RRAM is closely related to the tail-end thickness of the conducting filaments (CF) in the HRS conditions (Figs. 3b, 3c bottom panels). Because of the thicker Pt electrode edge compared to the graphene edge, the tail end of the CF will be thicker in the Pt-RRAM compared to the ones in the GS-RRAM. This greater thickness results in the more conductive HRS of Pt-RRAM.

The LRS of these devices are related not only to the size of the filaments but also to the different effects of oxygen in the TiN and the graphene electrodes. The LRS of Pt-RRAM (Fig. 2d) is comparable to that of GS-RRAM, even with larger filaments (Figs. 3b, 3c top panel). This is due to the effect of oxygen in TiN. It is fairly well known that oxygen forms a thin $TiO_xN_{1-x}$ film in the TiN layer, which works as a barrier against diffusion and carrier transport[24]. Such a barrier increases the interfacial resistance for Pt-RRAM ($R_{\mathrm{int,\ TiN}}$), and the total resistance at LRS becomes comparable to that of GS-RRAM.

The low SET/RESET voltage is related to the thickness of the electrode and the oxygen migration mechanism. After the forming process (Supplementary Figure 9), the tip of the conducting filament will be near the top electrode (Fig. 3c). The graphene serving as the SET electrode will have a much stronger electric field at the edge compared to the large TiN electrode because graphene is a monolayer thick. Therefore, a lower SET voltage will be sufficient to pull the oxygen ions from the oxide. On the other hand, we expect the lower RESET voltages are attributed to the lower activation energy for oxygen migration in graphene and the absence of a $TiO_xN_{1-x}$ diffusion barrier that is typically formed in TiN electrodes. The activation energy of diffusion for oxygen in graphene (0.15-0.8 eV, carrier density dependent)[25,26] is known to be lower than that of TiN (0.95-2.1 eV)[27]. Since the RESET mechanism is closely related to the oxygen diffusion assisted by Joule heating [20] and its activation energy, the required electrical potential for RESET will be lower for the graphene electrode than for the TiN electrode. The temperature-accelerated LRS retention-time measurement can probe the thermal activation of oxygen ion migration from the graphene to the oxide, as shown in Fig. 4b. From the linear fitting of the Arrhenius plot (Methods and Supplementary Figure 10), we estimate the activation energy for oxygen ion migration in graphene to be 0.92 eV, which is lower than the known values for TiN. It is worth noting that the work functions of graphene (4.56 eV) and TiN (4.5 eV) are comparable, and the difference in SET voltages cannot be explained by work function difference alone.

The result of the pulse mode endurance test in Fig. 4c indicated that the GS-RRAM maintained large memory window (> 70×) and showed no sign of deterioration after more than 1600 cycles of switching (Methods). The yield of the GS-RRAM (88%) was also comparable to that of the Pt-RRAM (92%). The reset current and the HRS/LRS characteristics of 10 randomly chosen GS-RRAM devices are shown in Fig. 4d. We also compare the 1st and the 2nd layer devices in Supplementary Figure 11.

### 3D array simulation

The storage density of a cross-point architecture is ultimately limited by the sneak-path leakage in the half-selected and unselected cells [16,28,29]. During the write operation, the extra voltage drop along the interconnects caused by the leakage current can lead to an insufficient voltage at the selected cell. During the read operation, parasitic conducting paths in unselected cells can degrade the output signal. To systematically investigate how the sneak-path leakage would limit the bit storage capacity of the 3D memory array, a Simulation Program with an Integrated Circuit Emphasis (HSPICE) circuit simulation [16,28,29] for the 3D array is performed, using the experimentally measured device properties (see Methods). Simulations are done using the worst-case data patterns[28] with the 0.5×$V$ write scheme and the column parallel read scheme [29]. The write margin ($V_{access}$ to the $V_{dd}$ ratio) and the readout margin ($\Delta I_{read}$ , the current difference between the on and the off state) as a function of total number of bits for the GS-RRAM and Pt-RRAM arrays are simulated under worst-case conditions assuming 200-layer stacks (Fig. 4e, f). The criteria that limit the total number of array bits during write and read operation are set at 70% and 100nA, respectively. In Figs. 4e and 4f, we observe that the write/read margin for GS-RRAM is larger and its degradation less pronounced, compared to those of Pt-RRAM, as the arrays become larger. This is a direct consequence of smaller pillar resistance enabled by thinner stacks of the graphene plane electrode with lower sheet resistance. Consequently, a larger array of graphene-based RRAM can be assembled without the adverse sneak-path leakage effect.

## Discussion

In this work, we demonstrated how the unique advantages of a 2D material can be exploited to outperform conventional materials in today's electronic applications. The E-field from the atomically thin edge electrode and the efficient ion storing/transport mechanism of graphene led to significantly lower power consumption. Graphene was also found to be the key enabler for ultra-high-density, bit-cost-effective 3D RRAM arrays. The increased density and the low power consumption of an RRAM structure will enable significant progress in emerging application areas such as energy-efficient abundant-data computing and neuromorphic computing[30]. RRAMs employing various oxides have already been demonstrated for spike-timing-dependent plasticity[30]. A highly integrated electronic synapse network employing low power graphene memory in a bit-cost-effective 3D architecture will be a significant step toward a highly efficient, next-generation computing system.

## Methods

### HR TEM sample preparation and imaging

The TEM-ready samples were prepared using the in situ FIB lift-out technique on an FEI Dual Beam FIB/SEM. For the imaging, we used an FEI Tecnai TF-20 FEG/TEM operated at 200kV in bright-field (BF) TEM mode or high-resolution (HR) TEM mode.

### Spatially resolved Raman spectroscopy

The images were taken with constant laser intensity right after the SET and the RESET programming. External perturbation was minimized with an oxide capping layer. For the purpose of Raman measurement, single-stack GS-RRAM (without the second stack) was fabricated and measured to eliminate any effect from the second graphene layer. A WiTec 500 AFM/micro-Raman Scanning Microscope was used for the 2D Raman raster scanning of graphene. A 532 nm wavelength was used for all measurements. A 30μm × 60μm area was scanned with an integration time of at least 4 seconds with a 1 μm resolution. Each measurement was conducted in less than 3 hours.

### Extraction of activation energy

The temperature-accelerated LRS retention-time measurement can probe the thermal activation of oxygen ion migration, as shown in Fig. 4b. This will cause the oxygen ions to migrate back to HfOx, increasing the resistance (i.e. RESET) of the RRAM. The kinetics of this process can be described by the Arrhenius law.

$$\tau_{reset} = \tau_0 \cdot e^{\frac{E_a}{k_B T}} \qquad (1)$$

The $\tau_{reset}$ is the characteristic time for RESET transition, $\tau_0$ is a constant, $k_B$ is the Boltzmann constant, $E_a$ is the activation energy barrier, and $T$ is the absolute temperature. The linear fitting result of retention time in logarithmic scale versus reciprocal temperature provides a good estimation of the activation energy (Supplementary Figure 10).

The measurements were done on a semi-automated probe system (Cascade Microtech, Summit) with a temperature controller (Temptronic SA166550). All measurements were done inside the test chamber with the nitrogen gas flowing. The setup was on an anti-vibration table with pneumatic vibration mount. The automated resistance measurement was conducted every 15 seconds to 3 min with 0.1 V bias using a semiconductor parameter analyzer (Agilent 4156C).

**Pulse mode endurance test**

The pulse mode endurance test was conducted with an Agilent Parameter Analyzer 4155C and an Agilent Pulse generator 81110A connected to a Keithley Switch Matrix 707B. Pulse width was 500 ns with 3s time delay and ±0.2V was the read voltage.

**HSPICE Simulations on the achievable array size**

We adopted the same resistance network and array simulation methodology for the worst-case selected cell of 3D RRAM as in ref [16,28,29]. The effect of the sneak-path leakage in the achievable array size can be quantified with the write margin ($V_{access} \times V_{dd}^{-1}$) and the readout margin ($\Delta I_{read}$). The definition of $V_{access}$ is the voltage across the accessed cell in the resistance network. $\Delta I_{read}$ is defined as the difference in the current flowing through the read resistor (100kΩ) when the RRAM cell is either in the HRS or the LRS. The HRS and the LRS values of GS-RRAM and Pt-RRAM were extracted from the experimental results of this work. $V_{dd}$, $V_{read}$, and $V_{half\text{-}bias}$ were set at 5V, 3.5V, and 2V, respectively. The maximum total bits for an array were determined using these criteria. The sheet resistance of Pt[15] and doped graphene[18] was assumed to be 300 Ω per square and 125 Ω per square, respectively. A selector parameter from a published result[31] was adopted for the simulation. The resistance of the selector was 57.9 MΩ at the half-bias condition and 1kΩ when it was turned on. During read programming, the selector was turned on and the resistance of the LRS of RRAMs was at least 5 times larger than the resistance of the selector during read operation. Feature size was 45 nm with a 12 nm selection material layer inserted in the pillar. The diameter of the Cu metal core was 5 nm and the thickness of TiN was 3 nm. The thickness of HfOx was 5 nm. Hence, the feature size was 2 × (5+3+12) + 5 = 45 nm.

**Additional information** Supplementary information is available in the online version of the paper. Reprints and permissions information is available online at www.nature.com/reprints. Correspondence and requests for materials should be addressed to S.L. (seansl@stanford.edu).

**Acknowledgements** This work is supported in part by the Office of the Director of National Intelligence (ODNI), Intelligence Advanced Research Projects Activity (IARPA) Trusted Integrated Circuits (TIC) Program (Program Directors: Dennis Polla and Carl McCants), and the member companies of Stanford Non-Volatile Memory Technology Research Initiative (NMTRI) affiliate program, and Systems on Nanoscale Information Fabrics (SONIC) Center, one of six centers of Semiconductor Technology Advanced Research Network (STARnet), a Semiconductor Research Corporation (SRC) program sponsored by Microelectronics Advanced Research Corporation (MARCO) and Defense Advanced Research Projects Agency (DARPA). J. Sohn is additionally supported by the STX foundation scholarship for overseas studies. Hong-Yu Chen is additionally supported by the Intel Ph.D. Fellowship.

Part of the work was conducted in the Stanford Nanofabrication Facility (SNF) at Stanford University, a member of the National Nanotechnology Infrastructure Network funded by the National Science Foundation.

**Author contributions** S.L., J.S. and H.S.P.W. conceived the experiments. S.L. and J.S. fabricated the devices, developed the electrical measurement set-up and performed the measurements. H.Y.C. provided support for fabrication. Z.J. performed the simulations based on experimental device properties. S.L., J.S. wrote the paper and H.S.P.W. supervised the work. All authors discussed the results and commented on the manuscript.

**Competing financial interests** The authors declare no competing financial interests

## Figure Captions

**Figure 1 Structure of graphene based and Pt based RRAM in a vertical 3D cross-point architecture.** **a,** An illustration of graphene-based RRAM in a vertical cross-point architecture. The RRAM cells are formed at the intersections of the TiN pillar electrode and the graphene plane electrode. The resistive switching HfOx layer surrounds the TiN pillar electrode and is also in contact with the graphene plane electrode. **b,** A schematic cross-section of the graphene-based RRAM. **c,** HR-TEM image (details in Methods) of the two-stack graphene RRAM structure. The RRAM memory elements are highlighted in red. The scale bar is 40nm. **d,e,** First and second layer of GS-RRAM with graphene on top of the $Al_2O_3$ layer. The scale bars are 5 nm. (f), (g), TEM image of the two-stack Pt based RRAM from previous work[14,15]. The scale bar is 40 nm for (f) and 5 nm for (g).

**Figure 2 The device characteristics of GS-RRAM compared to Pt-RRAM and other emerging memory devices.** **a,** Typical DC I-V switching characteristics of GS-RRAM and Pt-RRAM. For Pt-RRAM, SET process is observed when positive voltage is applied to TiN. For GS-RRAM, SET process is observed when positive voltage is applied to graphene. The SET compliances for G-mode, T-mode, and Pt-RRAM are 5 µA, 10 µA, and 80 µA, respectively for optimum conditions. A magnified plot of GS-RRAM is shown as inset. **b,** The SET and RESET voltage distribution of GS-RRAM and Pt-RRAM after 50 cycles of switching. The SET/RESET

voltages of GS-RRAM are noticeably lower (inset). **c,** Reset current distribution of GS-RRAM and Pt-RRAM after 50 cycles. GS-RRAM exhibit much lower reset current compared to Pt-RRAM. **d,** Resistance distribution after 50 cycles for GS-RRAM and Pt-RRAM at 0.1V. Larger memory windows are observed for GS-RRAM compared to Pt-RRAM. **e,** Reset power distribution of GS-RRAM and Pt-RRAM. The power consumption of GS-RRAM is 300 times lower than that of Pt-RRAM. This is from the combined effect of lower programming voltages and currents.

**Figure 3 The working mechanism and spatially resolved Raman imaging of oxygen ions in graphene during subsequent SET/RESET process of GS-RRAM**

**a,** Illustrations of the GS-RRAM structure. **b,** Working mechanism of Pt-RRAM. SET process (oxygen vacancy filament formation) is achieved by applying positive voltage to the TiN electrode. **c,** Working mechanism of GS-RRAM. The SET process is achieved by applying positive voltage to graphene instead of the TiN electrode. Notice the opposite direction of oxygen ion movement in GS-RRAM compared to Pt-RRAM. **d,** Changes in the 2D peak intensity as the oxygen is inserted (SET) and extracted (RESET) from the graphene film. The laser intensity was kept constant during the measurements. Notice that the reference silicon peak (520 $cm^{-1}$) is not changing during this transition. **e,** A microscopic image of the Raman mapped area highlighted in blue. The scale bar is 15 μm. **f,g,h,** 2-dimensional Raman scanning of the 2D peak intensity in the mapped area before programming (f), after oxygen ions are inserted into graphene via SET process (g), and after oxygen ions are pulled out from graphene via RESET process (h). All three images have the same color scale for 2D peak intensity and the laser intensity was kept constant during the measurements (See Methods). The darker hue is observed for graphene with the oxygen ions in (g). The scale bar is 10 um. The statistical distributions of the 2D peak intensity changes are also shown as histograms. Noticeable changes in the median values are observed as the oxygen ions are inserted into and pulled out from the graphene film.

## Figure 4 Resistance component breakdown, retention, pulse endurance, device variations, and array performance of stacked GS-RRAM.

**a,** Resistance component breakdown of GS-RRAM and Pt-RRAM. In comparison to Pt-RRAM, GS-RRAM has 4 different resistance components: Pt/graphene contract resistance ($R_{C}$), graphene film resistance ($R_{sh,G}$), graphene/HfOx interface resistance ($R_{int,G}$), and the thickness of the conduction filaments ($R_{Filament,G}$). **b,** Temporal evolution of GS-RRAM LRS resistance at temperatures ranging from 418K to 473K near 0.1V bias. Elevated temperatures were used in this study to obtain the critical time (i.e. filament rupture time) for oxygen migration within a reasonable time frame (See Methods). **c,** Pulse endurance test of GS-RRAM. Device switched with over 70× difference in HRS and LRS, and suffered no read/write disturbance after more than 1600 cycles. **d,** The maximum to minimum reset current distribution (top) and HRS/LRS resistance distribution after 50 cycles (bottom) for 10 randomly chosen GS-RRAMs. The cycle-to-cycle variations are shown as error bars which represent one standard deviation for each case. All devices were measured under the SET compliance current of 5 μA. The worst case scenario still exhibits HRS to LRS ratio exceeding 10×. **e,** Write margin comparison of Pt-RRAM and GS-RRAM for a 3D architecture with 200 stacks. **f,** Read margin comparison between Pt-RRAM and GS-RRAM for a 3D architecture with 200 stacks.

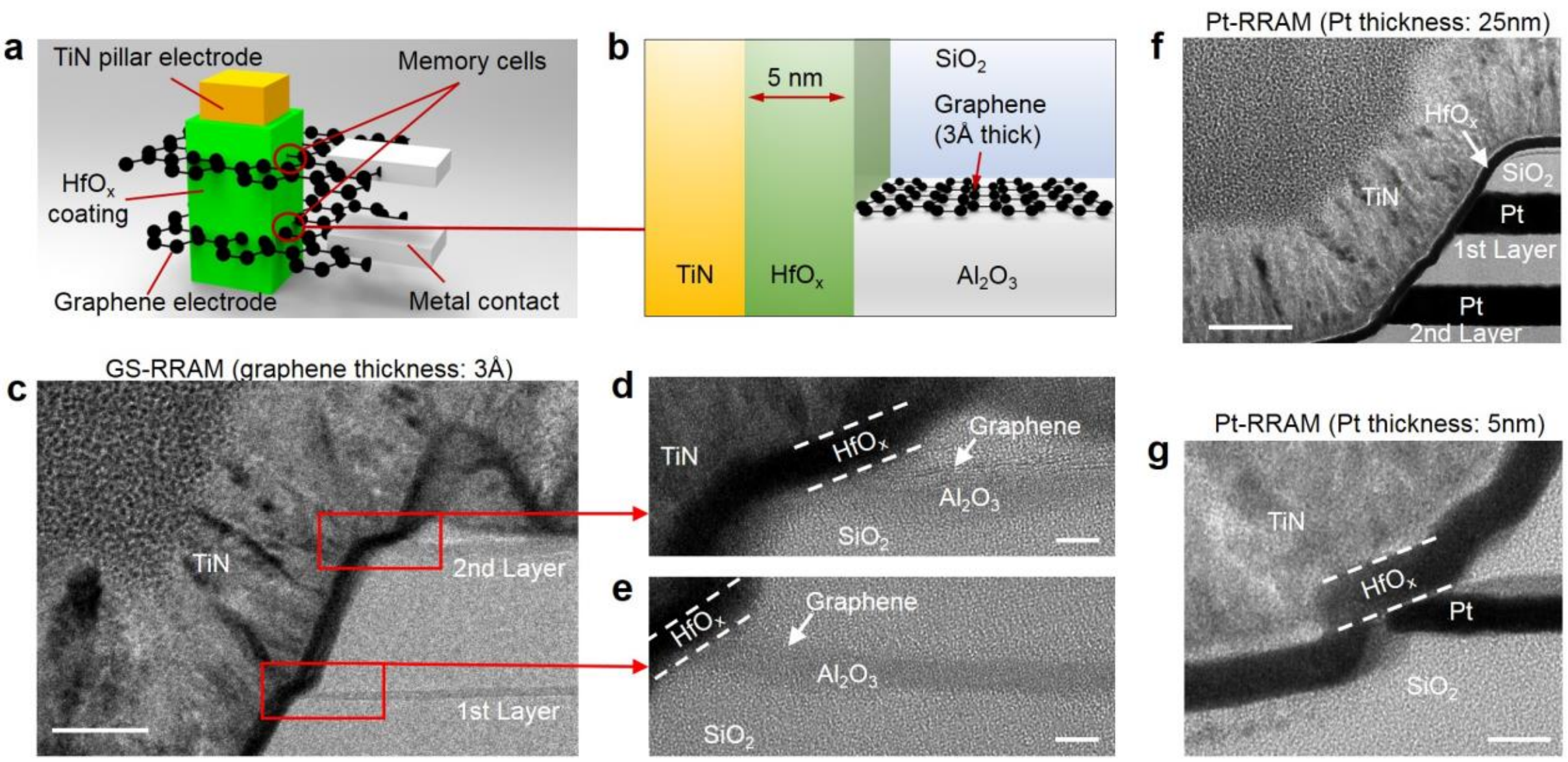

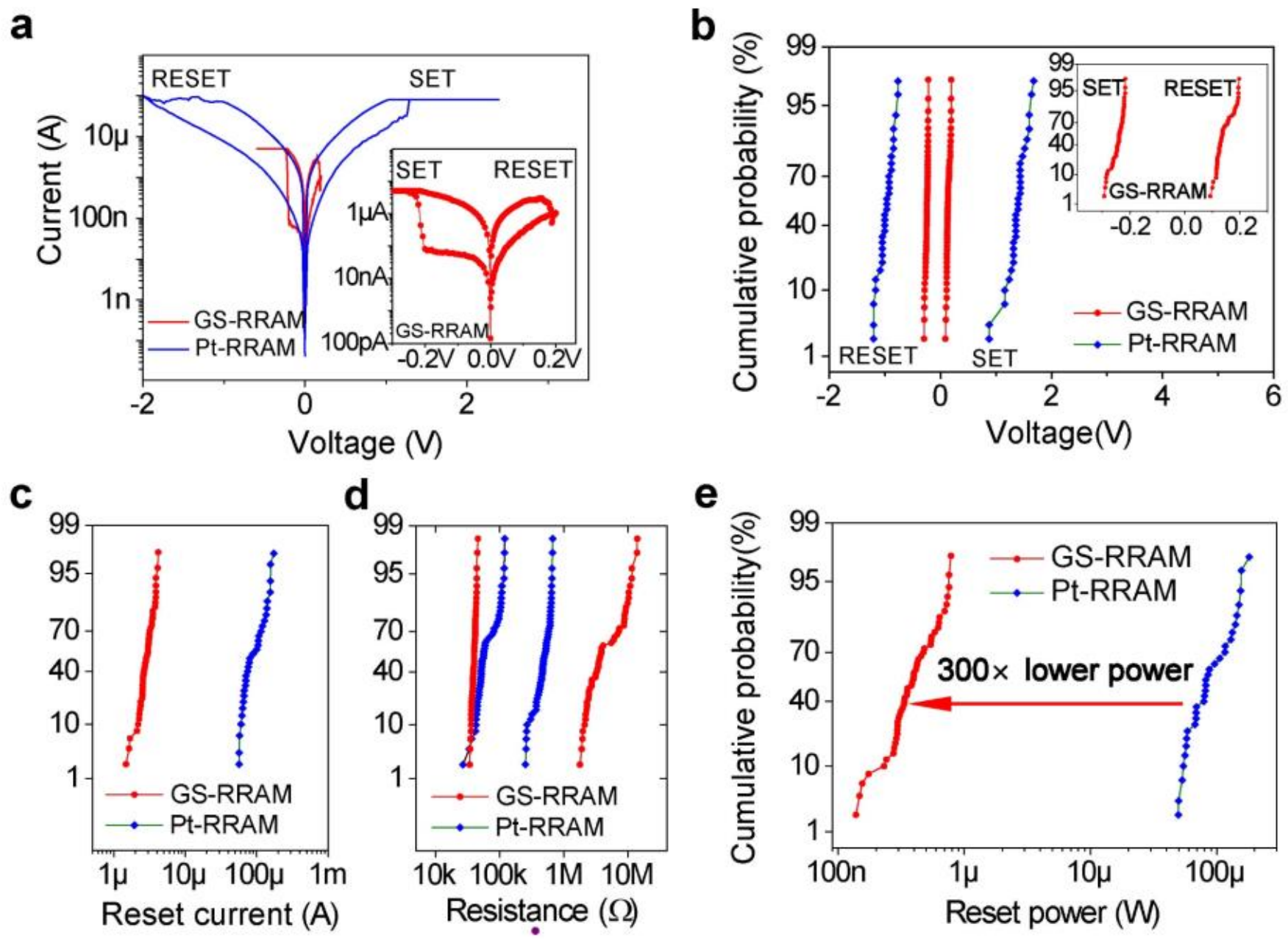

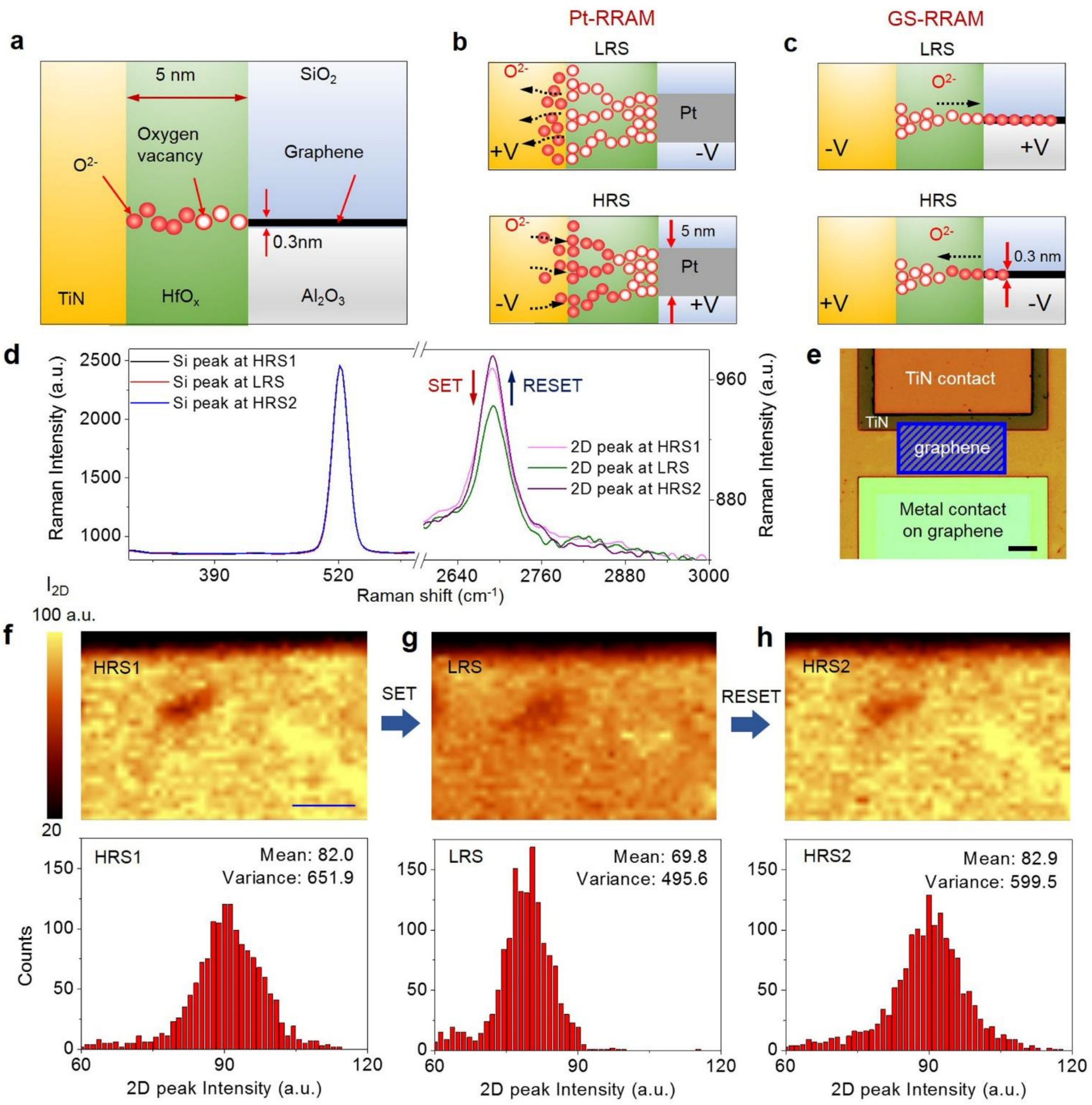

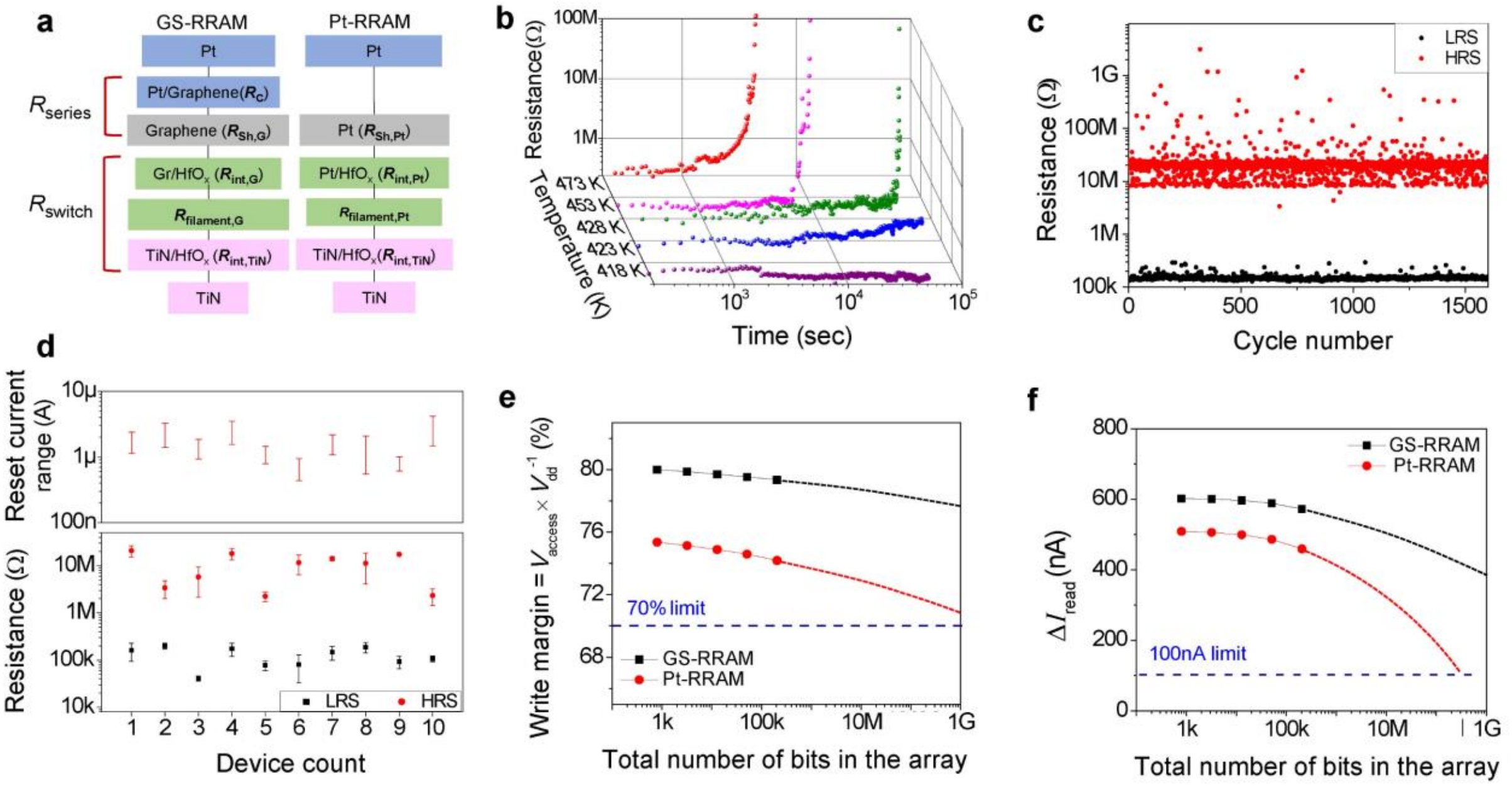

## Supplementary Figures

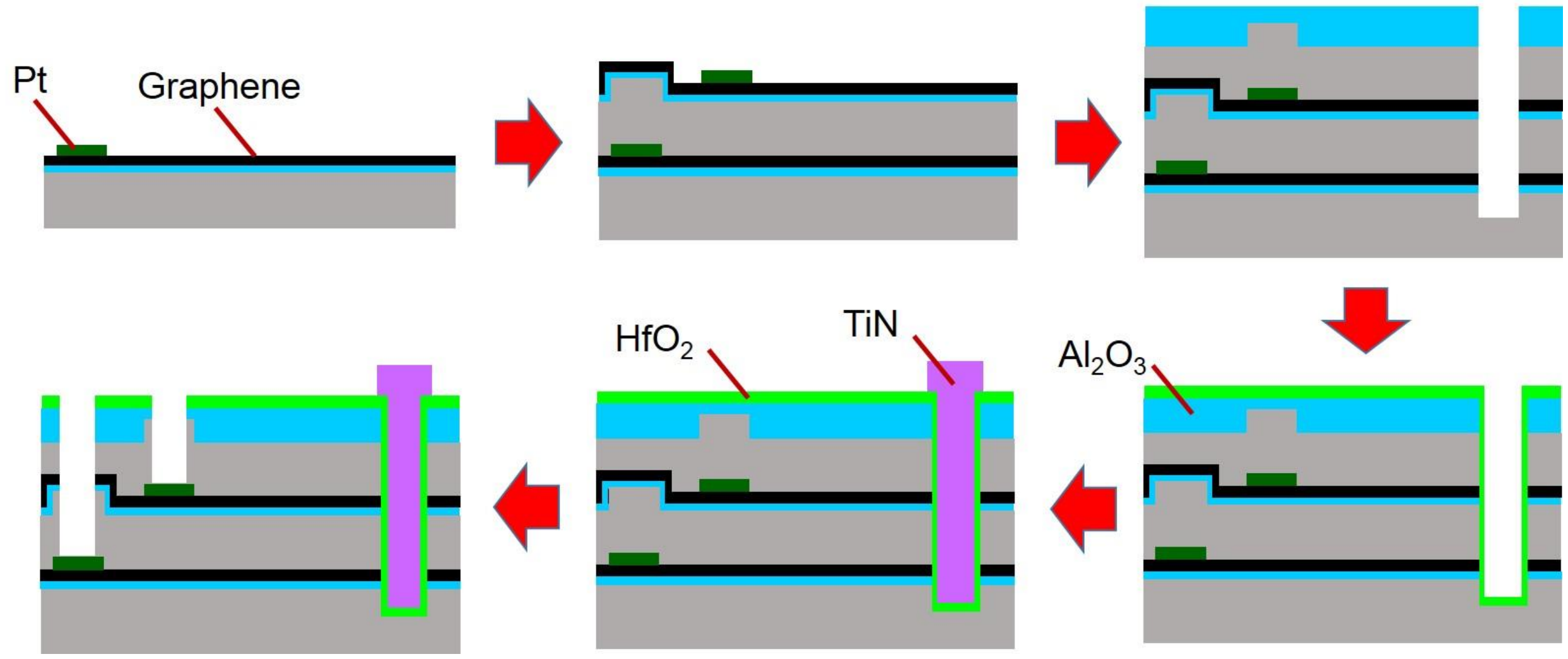

**Supplementary Figure 1 Fabrication process.** The devices were fabricated via sequential transfer of graphene, metal contact deposition, deep trench etching, $HfO_2$ deposition (atomic layer deposition) followed by TiN pillar formation (sputtering and lift-off) A thin layer (5nm) of $Al_2O_3$ was deposited before the graphene transfer process to promote graphene adhesion to the surface. First, single layer graphene is transferred on to a dielectric surface with 5 nm $Al_2O_3$ and 100 nm of $SiO_2$. The transfer method is identical to the previous works[1-3]. Monolayer graphene grown on copper foil with chemical vapour deposition method was purchased (Single Layer Graphene on Copper foil: 2 inch × 2inch, Graphene supermarket) and the monolayer quality was confirmed with Raman spectroscopy (Supplementary Figure 2c). Ti/Pt (3 nm/30 nm) layers are deposited by evaporation and patterned by lift-off process. 60 nm of $SiO_2$ (LPCVD) is deposited. Then these processes are repeated twice for two layers of single layer graphene; 50 nm ALD $Al_2O_3$ is deposited on the top layer for etch hard mask. A trench is etched down to the bottom $SiO_2$ layer followed by 5nm of $HfO_x$ (ALD) which is conformally deposited as the active resistive switching layer and 200 nm of TiN electrode is deposited by sputtering and patterned via lift-off. The contacts are opened via dry etching.

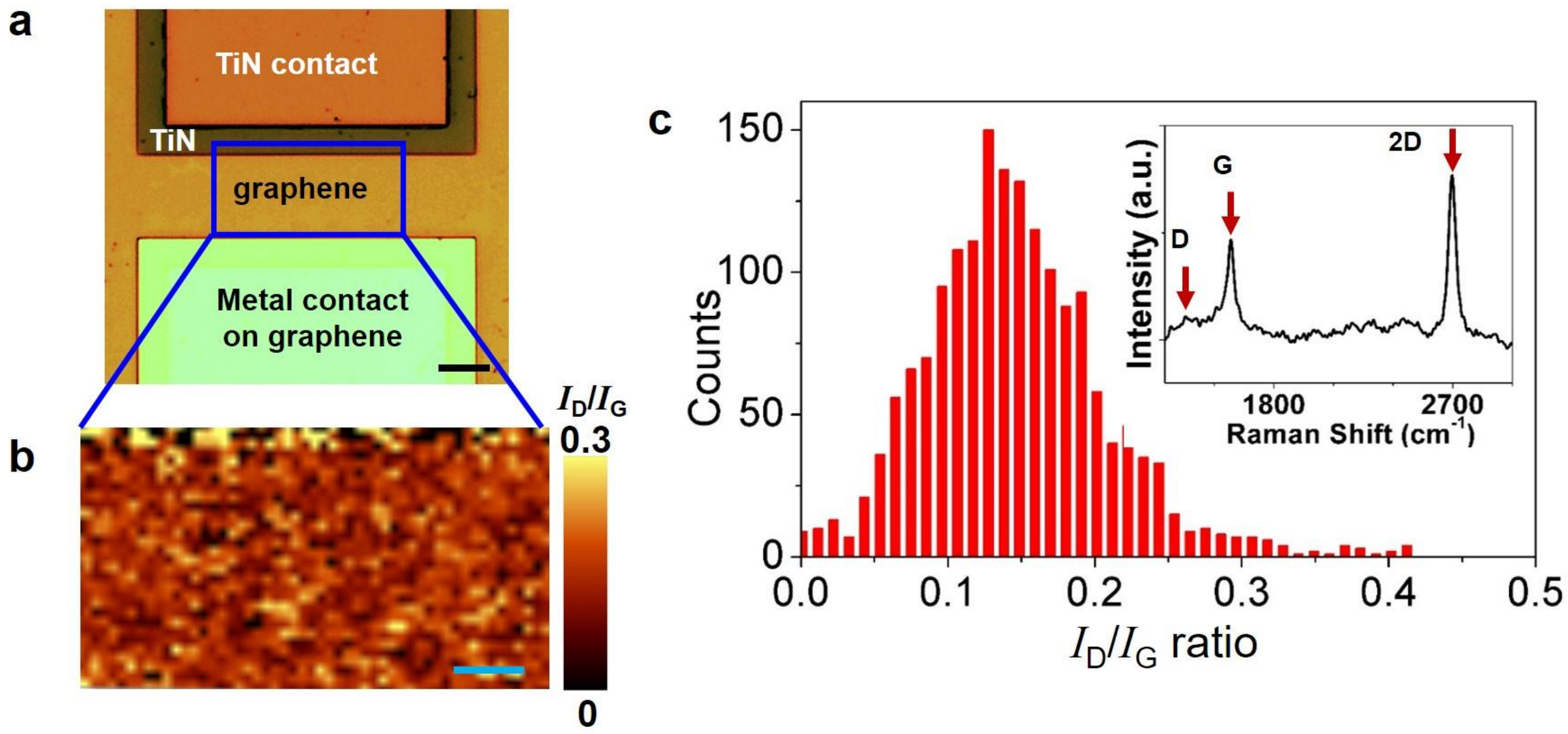

**Supplementary Figure 2 Verification of graphene thickness and quality. a,** Actual device image. The Raman laser scanned area is highlighted in blue. The scale bar is 15 µm. **b,** A 2D Raman spectra map of D peak to G peak ratio ($I_D/I_G$) after the complete fabrication process. This ratio is a known indicator of the disorders in graphene films. The $I_D/I_G$ value is limited to approximately 0.1, indicating a low defect density in the film[4]. The inset shows a typical Raman spectrum of monolayer graphene with weak D-peak intensity after the complete fabrication process. Minimized physical disturbance and the low fabrication temperature (<300 ℃) were essential to maintaining the high quality graphene. The scale bar is 10µm. **c,** A histogram of $I_D/I_G$ ratio of Supplementary Figure 2b. The median value is 0.12. A typical Raman spectrum of the scanned area is shown as the inset.

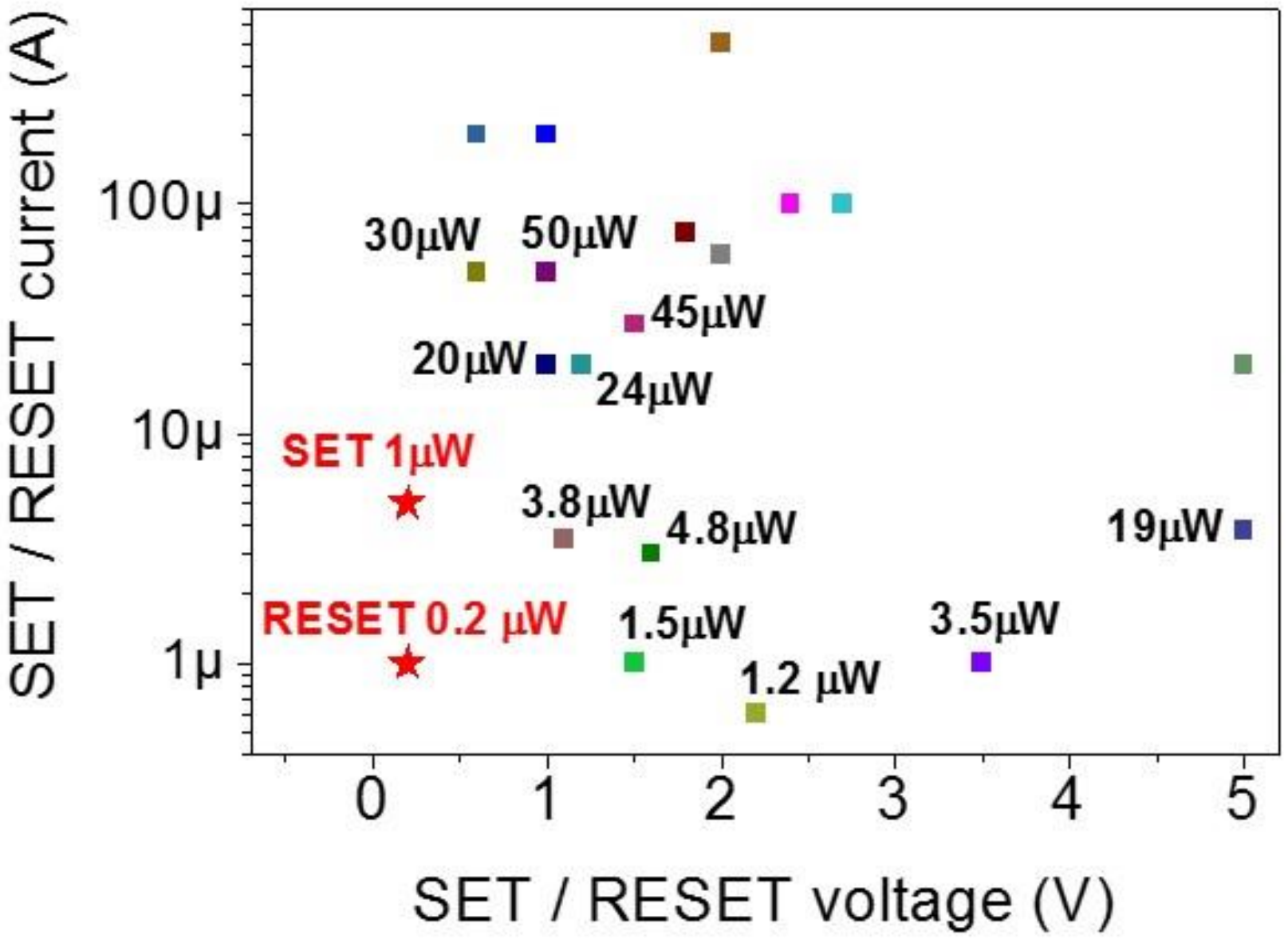

**Supplementary Figure 3 Comparison of power consumption.** Programming voltages, currents, and power consumptions from the recent reports[5-23] on low power RRAMs were plotted. With one of the lowest SET/RESET voltages ever recorded, the SET and the RESET power consumption of the demonstrated GS-RRAM (shown as red stars above) exhibit extremely low values. From a practical application point of view, the process that consumes the most power (SET or RESET) is plotted for other works, since the larger value determines the power delivery requirements for the chip.

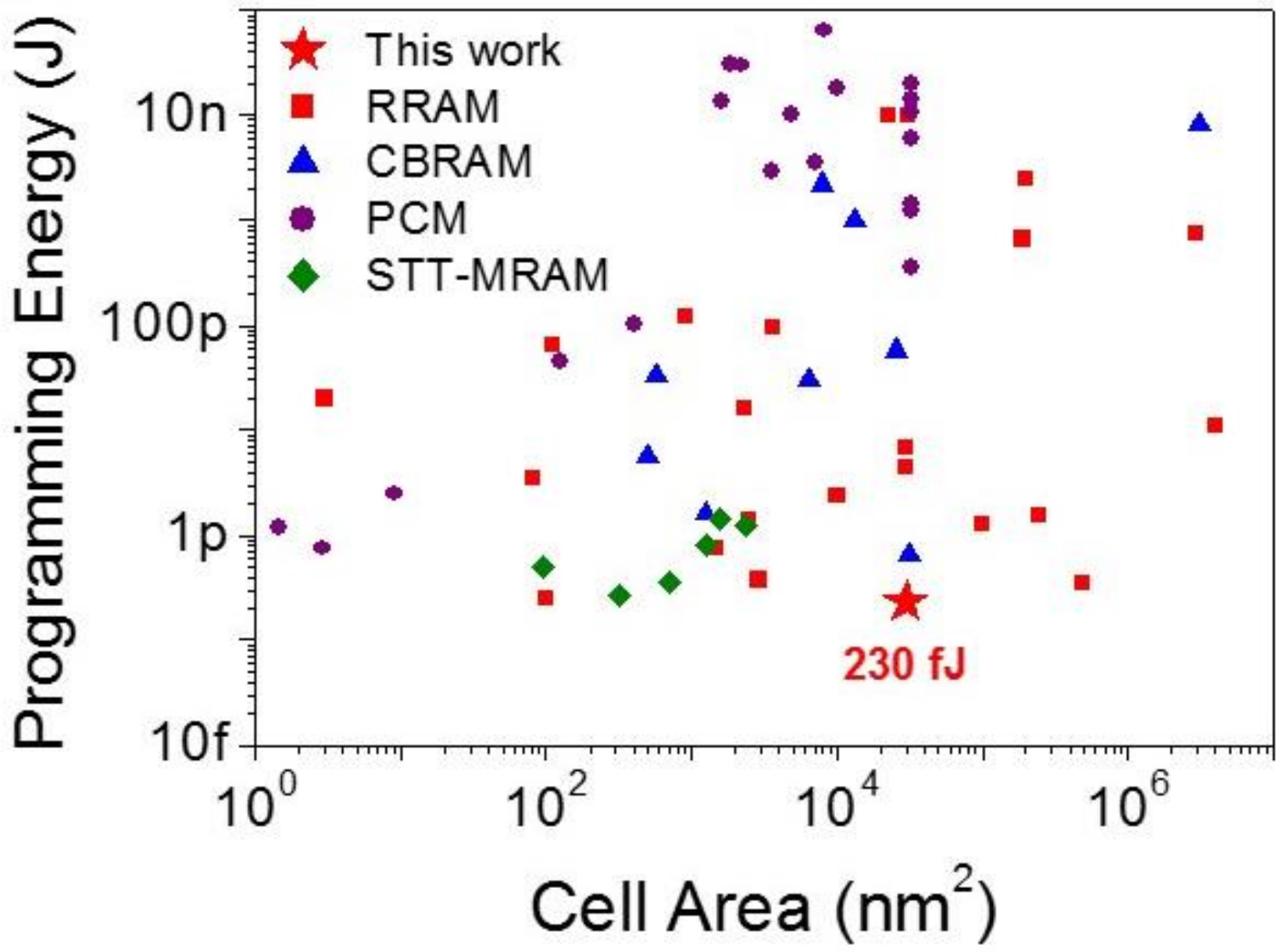

**Supplementary Figure 4 Comparison of energy consumption.** Comparison of programming energy for GS-RRAM and other emerging non-volatile memories with respect to cell area. The switching energy for GS-RRAM is one of the lowest. RRAM references are [5,11,24-47], CBRAM references are [48-56], PCM references are [57-76], and STT-MRAM references are[77,78], repectively.

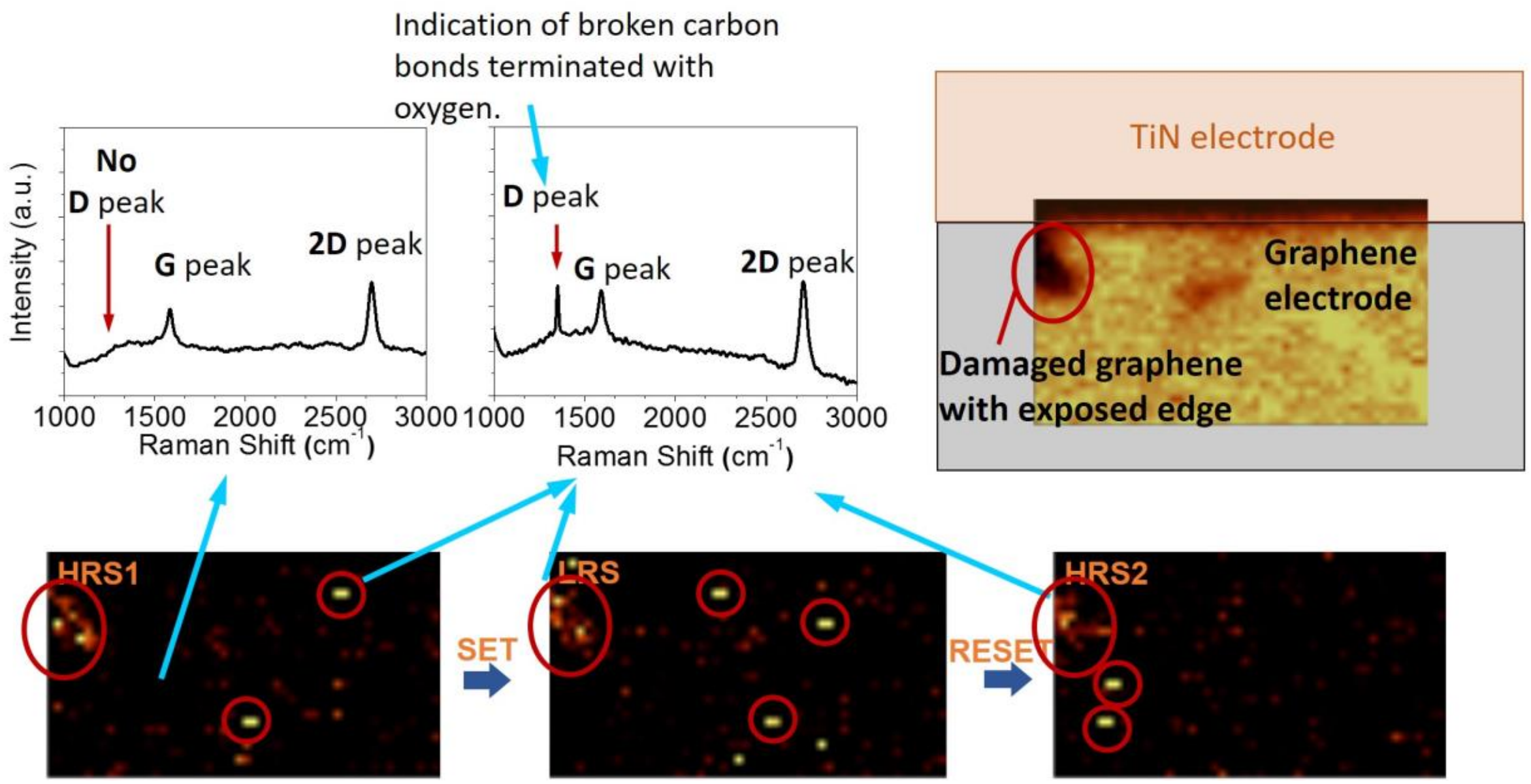

**Supplementary Figure 5 Oxygen bonding in the graphene electrode.** Although graphene is widely known to be inert, the edge and the broken bonds at the defect sites are more active compared to the basal plane of the graphene sheet. Typical graphene oxide Raman signature is the pronounced D peak[79]. (The intensity of G peak, on the other hand, is associated with the number of graphene layers and this may or may not be related to the graphene oxide.) D peak is also a strong indication of broken carbon bonds (i.e. dislocations, defects) and is pronounced in graphene ribbons with the edges exposed. These broken carbon bonds are more likely to be terminated with oxygen atoms. We specifically found an area in one of devices where the graphene was damaged and the edge was exposed. This edge is composed of broken carbon bonds similar to defects/dislocations at the basal plane and can be detected with the D peak intensity map as shown below in Supplementary Figure 3. An interesting aspect is that the defect region (bright area) highlighted with red circles seems to be created/annihilated (or even shifted) after consecutive SET and RESET process. Several past research results confirm that graphene broken bonds (dislocations) can be created/annihilated and shifted depending on which state is more thermodynamically favorable[80,81]. More importantly, this indicate that these oxygen binding phenomenon is reversible as previous work[82] suggested. As indicated in the reference[82], the oxygen may form a covalent bond at the defect sites of graphene after the SET process and the process is reversed during the RESET process. Another important observation is that the point defects seems to be created and annihilated randomly but at the edge, the bright colored region is pervasive regardless of whether it is after the SET or the RESET process. This may indicate that the edge is always oxidized when it is in contact with the $HfO_x$. The oxidized edge seems to have little effect on switching endurance of the device. We have switched the device more than 1600 times to observe that the memory function did not degrade (Fig. 4c).

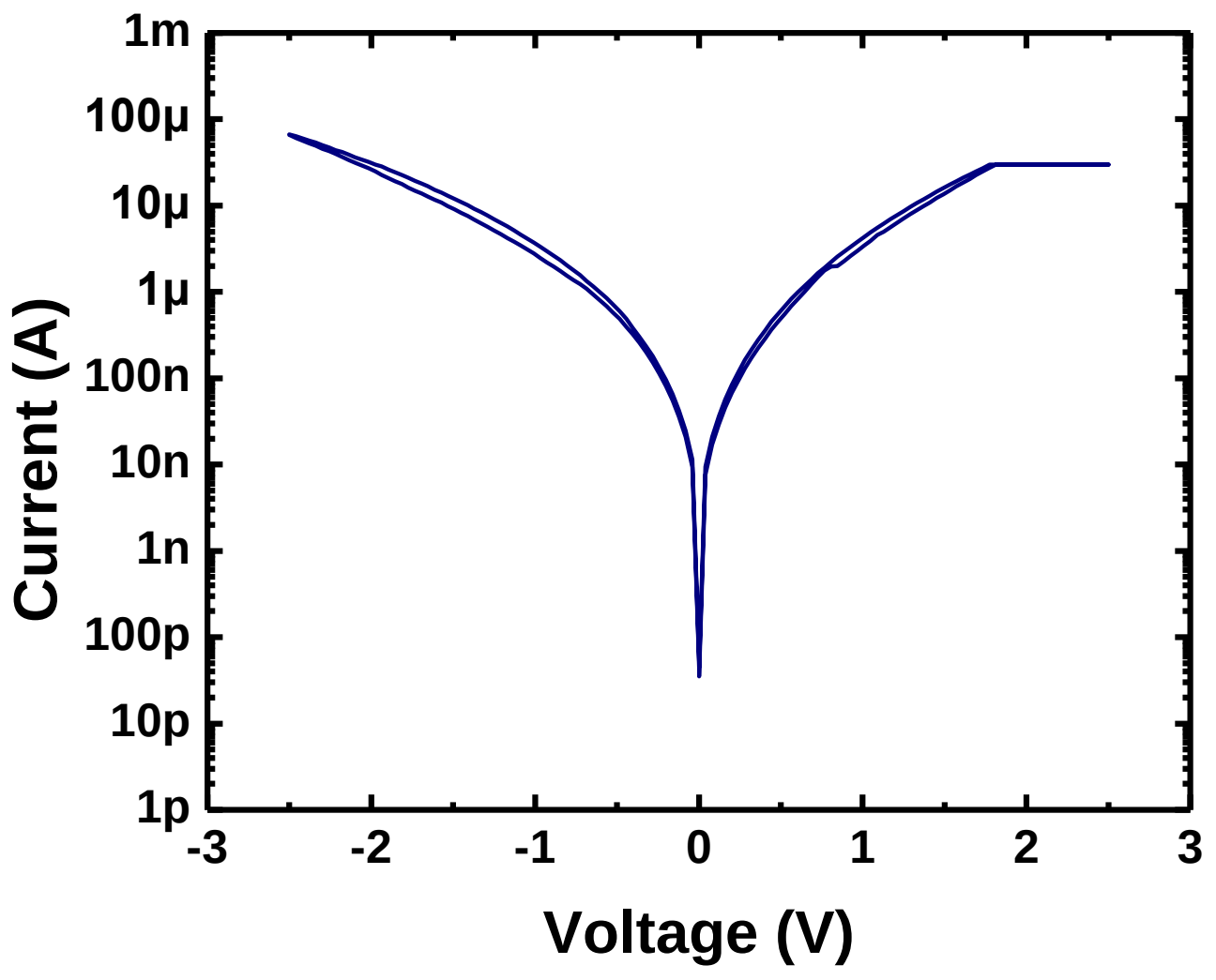

**Supplementary Figure 6 Degradation of memory window for Pt-RRAM with 30 μA SET compliance.** Pt-RRAM devices with lower SET compliance than 80 µA suffers from memory window degradation as shown in the plot. This is expected since PtRRAM's HRS is significantly more conductive compared to GRRAM due to the larger area of the Pt bottom (passive) electrode.

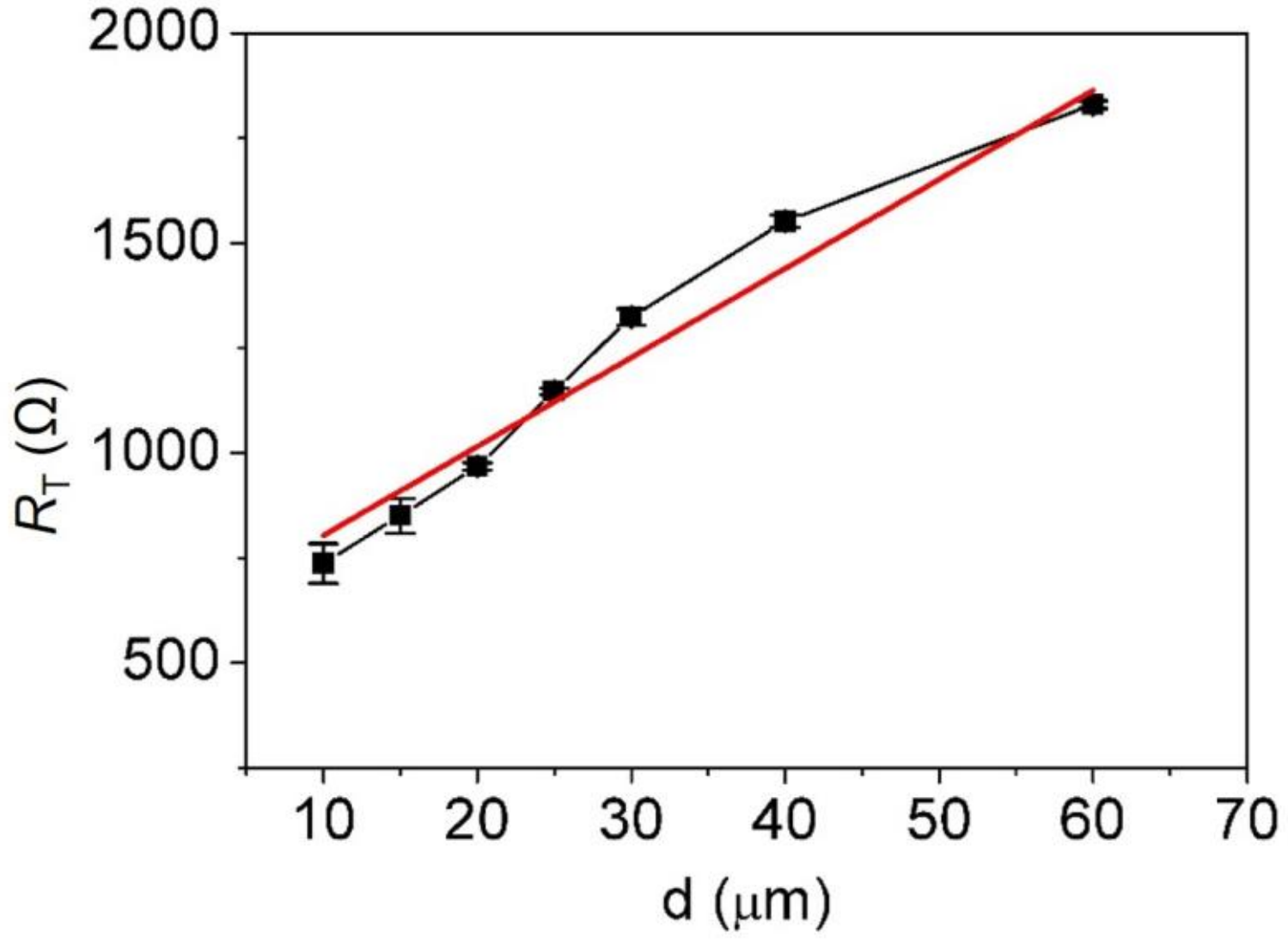

**Supplementary Figure 7 Total resistance value from the as-fabricated wafer with circular transmission line test structure[83] as a function of gap distance.** From the Y-intercept = $2R_C$ = 591Ω. The corresponding contact resistance $R_C$ between the graphene and the metal (Ti/Pt) contact was found to be 295Ω with specific contact resistance of 9.3Ω·cm. With the slope of 21.2 Ω $\mu m^{-1}$, the sheet resistance of graphene ($R_{sh,G}$) is extracted to be 6.7kΩ per square. Pristine, exfoliated graphene without environmental doping is reported to have sheet resistance value of ~6 kΩ per square.[84]. From our $I_D/I_G$ Raman map (Supplementary Figure 2), the defect level was not significant after dielectric deposition (LTO, 300℃). Considering the low D-peak level in our graphene, the resulting $R_{sh,G}$ is in close agreement with that of a pristine graphene that is void of any dopants or defects [85-87]. Since the measurements are done on the as-fabricated wafers, small discrepancies may arise from the process conditions. For Ti/Pt layer (Ti 1 nm/Pt 5 nm), the sheet resistance $R_{sh,Pt}$ was extracted to be 558Ω from 20 TLM measurements. The Pt sheet resistance is also in agreement with the literature [88]. Graphene is approximately ×20 thinner than the Ti/Pt layer and ×12 more resistive, showing slightly superior conductance with similar thicknesses. However, it should be noted that atomically thin metal such as Pt layer tends to form discontinuous island, and a sharp nonlinear increase in sheet resistance is observed as the thickness decreases [88].

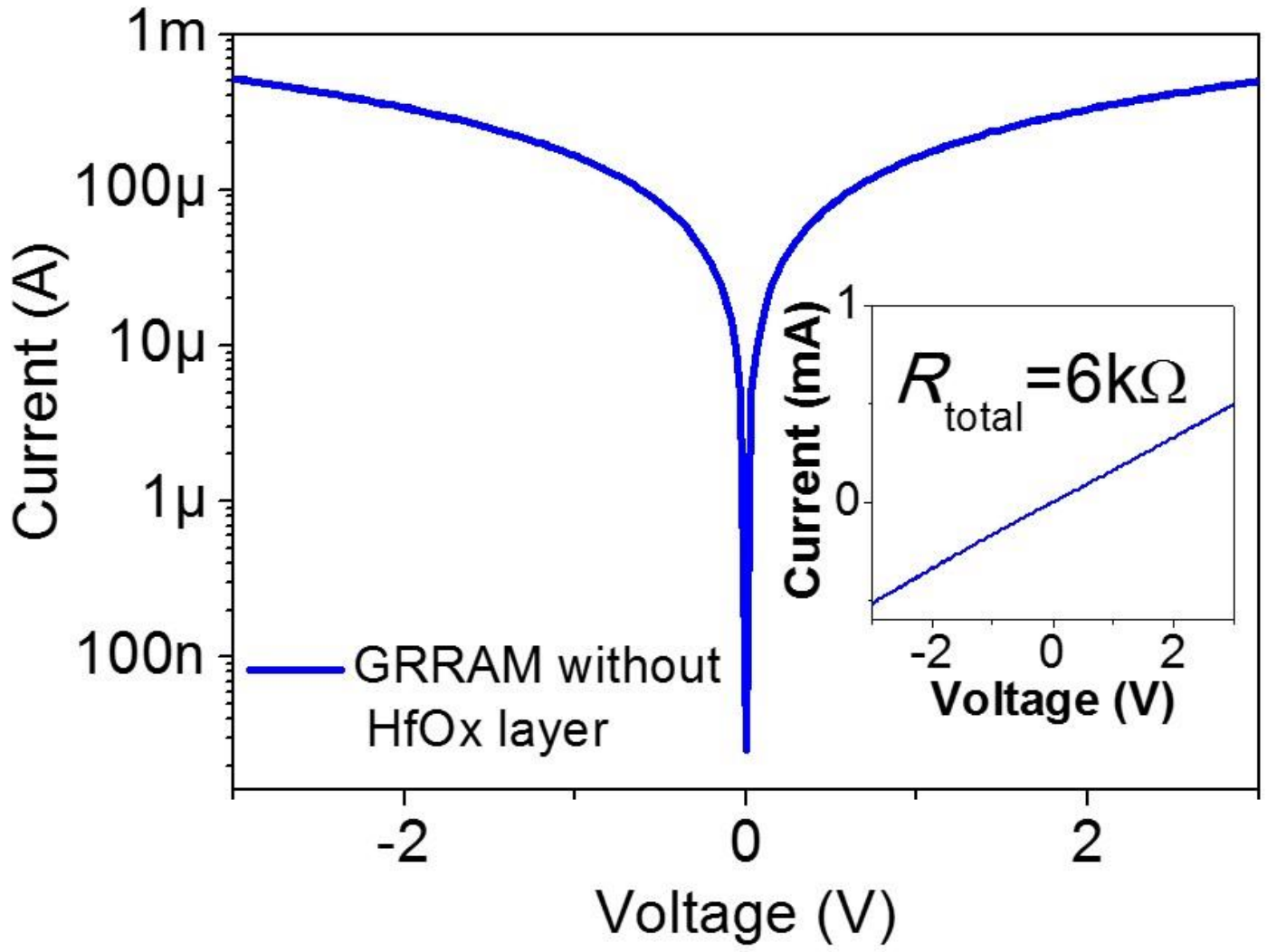

**Supplementary Figure 8 The I-V curve of GS-RRAM without the HfOx layer (inset: linear scale).** The I-V curve of GS-RRAM without the HfOx layer (inset: linear scale). The total resistance of the GS-RRAM device *without* the $HfO_2$ is close to 6 kΩ, which is only a fraction of HRS resistance. This strongly indicates that the series resistance $R_{series}$ (i.e. $R_{sh,G}$ + $R_c$) of GS-RRAM is not the major factor that contributes to the increases of the HRS resistance in GS-RRAM. On the contrary, this outcome suggests that the difference between the $R_{switch}$ of GS-RRAM ($R_{int,G}$ + $R_{filament,G}$) and Pt-RRAM ($R_{int,Pt}$ + $R_{filament,Pt}$) determines the HRS of GS-RRAM and Pt-RRAM, respectively.

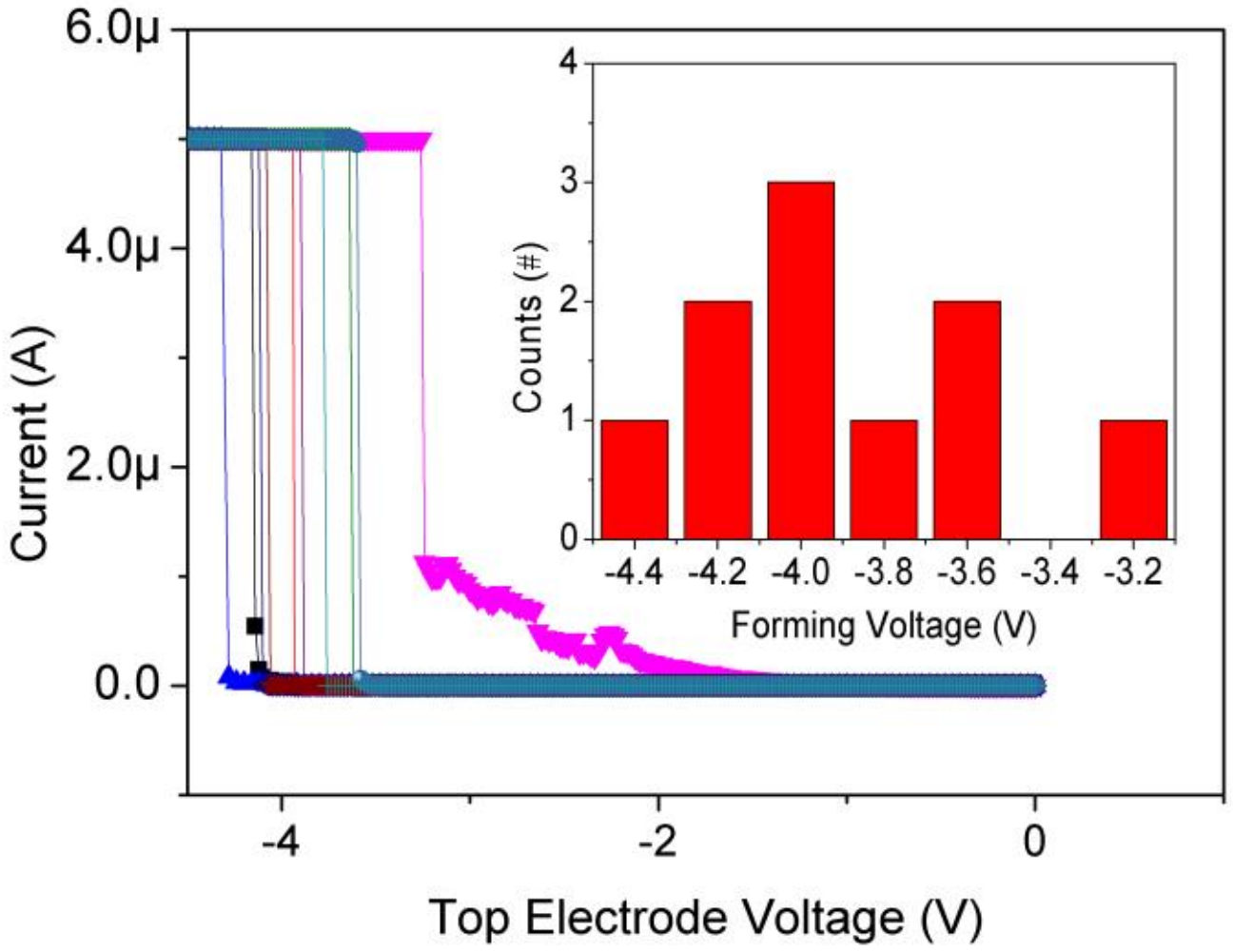

**Supplementary Figure 9 Forming of GS-RRAM.** The top electrode is the TiN electrode during the forming process. Forming curves are collected from 10 cells with 5µA compliance current. Inset: forming voltage distribution.

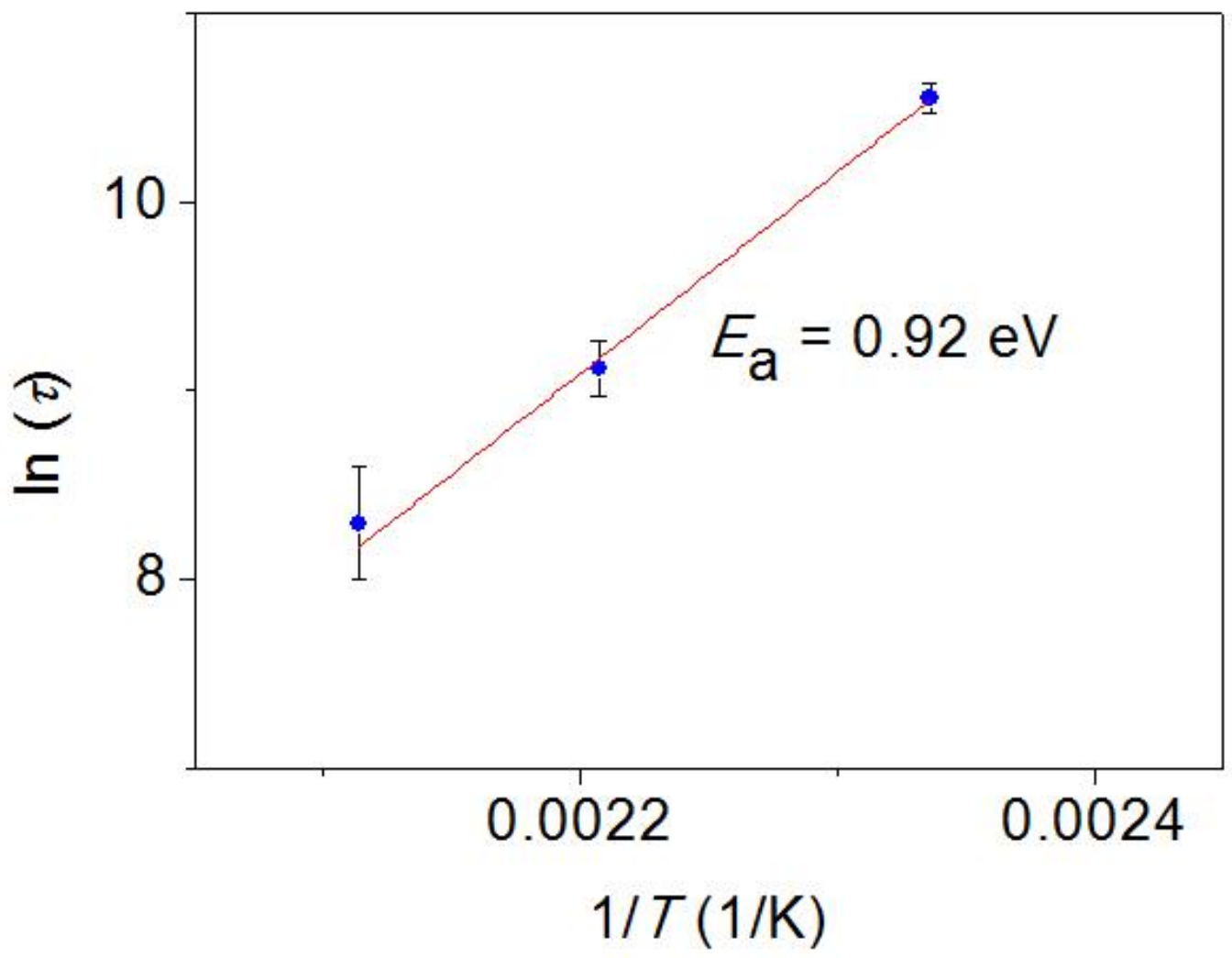

**Supplementary Figure 10 Arrhenius type plot of the wait time versus 1/*T* extracted from Fig. 4b.** The sudden transitions to the OFF state in Fig. 4b corresponds to the rupture of the oxygen vacancy based filament by the diffusion of oxygen ion towards $HfO_x$ layer. The activation energy of the barriers can be extracted from temperature dependence of the characteristic dwell time for RESET transition (Arrhenius equation in Methods). From the linear fitting of retention time in logarithm scale versus reciprocal temperature, we estimate of the activation energy $E_a$ for ion migration in graphene to be 0.92 eV.

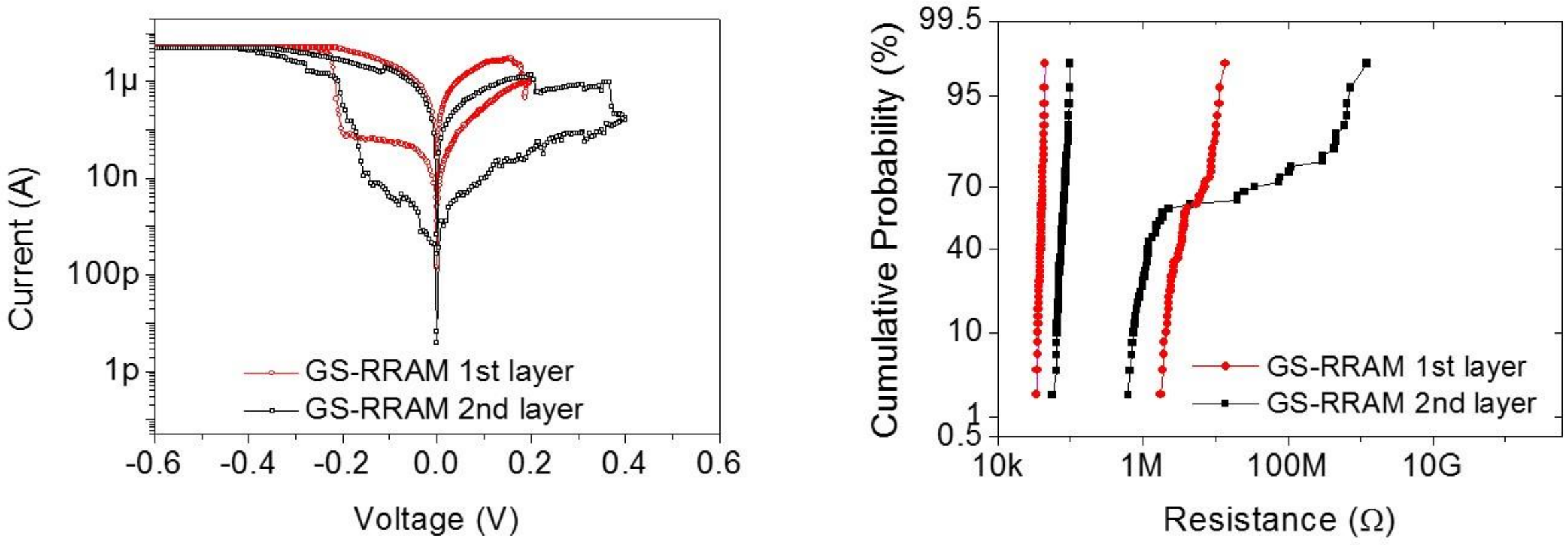

**Supplementary Figure 11 Typical DC I-V switching and HRS/LRS characteristics of bottom layer and top layer of GS-RRAM.** The GS-RRAM in the bottom layer exhibited even lower RESET current with similar SET voltages. However, there were some discrepancies in the RESET voltages for bottom and top layer. Importantly, the overall RESET power is still similar due to lower RESET current. Qualitatively similar memory windows were observed for top and bottom devices. The lowest memory window in the second layer is still above 10×.

## Supplementary Table

| | Thickness in nm (Electrode+$SiO_2$) | *F* = 20 nm | *F* = 22 nm | *F* = 26 nm |
|---|---|---|---|---|
| Etch ratio 30<br>Etch Angle 88° | 11nm (Pt=5nm) | 54 | 60 | 70 |
| | 16nm (Pt=10nm) | 37 | 41 | 48 |
| | 21nm (Pt=15nm) | 28 | 31 | 37 |
| Etch Angle 89° | 6.3nm (graphene=0.3nm) | 181 | 200 | 236 |

**Supplementary Table 1 Analysis of the number of achievable stacks with a dielectric thickness of 6nm**. The achievable number of stacks can be calculated using the equation for reliability projection from reference [88]. Total stack height = $R \times F/T$ ($R$ is the etching aspect ratio, $F$ is the lithographic half pitch, and $T$ is the combined thickness of the plane electrode and the dielectric in between). Assuming $SiO_2$ thickness of 6 nm, half-pitch of 22nm, and etch angle increase of just 1°, the maximum graphene RRAM stacks possible will be 200 stacks compared to the 60 stacks possible with Pt-RRAM. With an operating voltage of 0.2V in our GS-RRAM and a higher etching angle, we expect the number of possible graphene RRAM stacks to increase even more since a thinner dielectric can be used.

# Supplementary Notes

## Supplementary Note 1 3D vertical cross-point architectures

A pressing imperative for RRAM technology is to adopt a bit-cost-effective 3D architecture satisfying the requirements of performance metrics (density, latency, and energy consumption), which surpass those of 3D stackable multi-bit NAND Flash technology. Many industry/research groups [89-93] are actively working on variations of 3D vertical cross-point architectures as shown in Supplementary Fig. S1. The graphene RRAM in this work (with pillar electrode and planar graphene electrode) is compatible with all the 3D vertical cross-point architectures recently introduced [89-93].

The integration density of such 3D architectures depends on the number of stacks which is limited by the plane electrode thickness, the sheet resistance of the plane electrode, the dielectric thickness (related to the programming voltages and cross-talk), the pillar etch angle, the lithographic pitch, and the resistance of the pillar/plane electrode [88,94].

Since the total pillar height is limited, a thin device structure will be important for ultra-high density storage[88,94]. However, there is a fundamental limitation on how thin the metal plane electrode can be.

There has been a recent report of an RRAM structure with a sub-5nm thick vertical TiN electrode [95]. Although it is possible to form such sub-5nm metal electrodes, the main challenge lies not in the thickness of the metal, but in the high sheet resistance. All metal films are known to exhibit a steep exponential increase in sheet resistance as the thickness decreases under 10 nm[88,96]. This is because extremely thin metal films tend to form discontinuous islands, and thin dielectric layers are formed on the grain boundaries[96]. Such high sheet resistance of the plane electrode will result in a significant voltage drop on the electrode and severely degrade the write/read margin of the 3D RRAM structure[88,94], which limits the integration density. Hence, producing a sub-5nm conducting film with a low enough sheet resistance for 3D RRAM is a difficult task without using special methods or materials.

Graphene's sheet resistance per thickness is significantly lower than that of any metal. Graphene has been experimentally proven through the use of doping technique[97] to have sheet resistance as low as 125 - 200 Ω per square[1,86,97] with a monolayer thickness. These levels of resistance are something impossible to achieve (at such thickness) with conventional metal. From the measurements, graphene exhibited superior sheet resistance value per thickness (i.e. graphene is 20× thinner and 12× more resistive) compared to Pt after fabrication (Supplementary Section 7). Considering the nonlinear increase of Pt sheet resistance in such a scale, the actual sheet resistance of Pt when it is as thin as graphene will be drastically higher.

It is also important to note that metal contact to graphene is an ohmic contact, and the contact resistance is relatively low due to the graphene's semi-metallic nature[97]. An optimized metal/graphene specific contact resistivity is $7.5\times10^{-8}$ Ω $cm^2$ [98]. This value is smaller than that of both Al and Pt contact to degenerately N-doped silicon ($2\times10^{20}cm^{-3}$) as shown in [99].

From the analysis in the previous work[88], the required dielectric thickness is approximately 6 nm of $SiO_2$ in between each layer if the devices are to work with operating voltages of 3V (much higher than the 0.2V required for our GS-RRAM). The 6 nm $SiO_2$ is required since it can maintain a lifetime > 10 years at the operating voltage of 3V based on the breakdown voltage and the time dependent dielectric breakdown (TDDB) lifetime extrapolation for PECVD $SiO_2$ sandwiched between metal electrodes[100]. Finally, graphene (3Å ) is significantly easier to etch vertically than Pt (6nm) during pillar formation. (Graphene is simply etched with weak $O_2$ plasma treatment.) This property is highly beneficial since the etch angle is a very important factor that determines the number of achievable stacks [88,94].

## Supplementary Note 2 Comparison of using graphene as an oxygen detector (previous work, ref[82]) and for oxygen storage (this work)

In RRAM devices, the resistive switching is attributed to the formation (SET) and the subsequent rupture (RESET) of nanoscale conductive filaments involving oxygen ion migration[14,101-106]. A generally accepted theory claims that the filament formation is based on the oxygen ion movement from the switching material.

It is fairly well known that the oxygen function as dopants in graphene, and the doping level of graphene can be observed with Raman spectroscopy[82,107,108]. We have previously monitored the oxygen ion in a RRAM structure by inserting graphene film between the TiN layer and $HfO_x$[82].

The memory structures in our previous work and the current work are very different. In the previous work, the SET electrode is the TiN and the RESET electrode is the Pt. In our work, the SET electrode is the graphene edge and the RESET electrode is the TiN. Also the previous work is a planar structure and the current work is a vertical structure.

Although both previous and current work report low power consumption, the mechanisms for achieving low power consumption are fundamentally different. In the previous work, the low power was due to reduced RESET current from the high built-in series resistance of inter-layer graphene. The overall SET/RESET voltages (~2V) have few differences between structures "with" and "without" graphene interlayer.

In the current work, we see a drastic difference in SET/RESET voltages between GS-RRAM (~0.2V) and the Pt-based device (~1.5 to 2V). This is because the graphene, instead of the TiN layer, is used as the SET electrode. Here we are using graphene as a stand-alone oxygen reservoir, unlike in the previous work. The lowering of SET/RESET voltage is related to the lack of a $TiO_xN_{1-x}$ barrier layer in the HfOx/graphene interface, and the ease of oxygen diffusion across the graphene electrode as explained in the main text.

# Supplementary References